**Mathematical theory of diffusion in solids: solutions in the semi-infinite body and solution to a diffusion problem with a variable boundary condition.**

**Guglielmo Macrelli (*)**

(*) Isoclima SpA – R&D Dept. Via A.Volta 14, 35042 Este (PD) Italy
guglielmomacrelli@hotmail.com

**Abstract:**

A review of solutions of solid-state diffusion problems in infinite and semi-infinite bodies is presented. Based on the identified solutions for the semi-infinite body a two-step diffusion problem is discussed in detail with the first step characterized by a Dirichlet constant concentration condition and the second step by a Neumann condition.

## I. Introduction

In solid state diffusion problems it is common to have diffusing elements in a solid body that can be externally injected from deposited layers, coming from other phases either liquid or gaseous or internally moving or redistributed from internal locations in the solid matrix. When the diffusion length $\lambda_D$ is very small compared to the sample dimension (usually the thickness $d$) through which diffusion occurs $(\lambda_D << d)$ than we can approximate the solid to a semi-infinite half body. The main purpose of this study is to provide a detailed proof and justification for a variable boundary condition problem which solution has been provided by the author in the literature[1,2] without a proof. The reason for not providing the proof was mainly related to the length and complexity of the proof itself. There can be found mentions of this problem in the literature[3,4] that for different reasons do not fully cover the purpose of this study. The solution provided by Malkovich[3] inspired the approach used in this study but it is limited to a situation where the two steps diffusion processes are run with the same diffusion coefficient (same temperatures). The solution provided by Kennedy and Murley[4] is according to the Green functions approach but it contains presumably a typo in the final solution. In this study we have considered worth to preliminary review the matter of solutions to diffusion problems in the infinite and semi-infinite bodies as a necessary step before approaching the more complex variable boundary diffusion problem. We consider worth to have provided a detailed proof of this solution first to cover a missing gap second because the constructional approach to the proof may be beneficial for similar problems in solid state diffusion. This is a contribution to the mathematical theory of solid state diffusion covering some gaps of past literature.

## II. Solution in the infinite and semi-infinite body

In this study we deal with the differential equation for the concentration of diffusing elements into a solid body. We consider the mono-dimensional version of the problem. The starting point is the Fick equation for the flux $J$ of the diffusing elements defining a diffusion coefficient $D$:

$$-J(x,t) = D \frac{\partial c(x,t)}{\partial x} . \qquad (1)$$

Together with the continuity equation:

$$\frac{\partial c(x,t)}{\partial t} + \frac{\partial J(x,t)}{\partial x} = 0 \qquad , \qquad (2)$$

it results the diffusion equation for concentration:

$$\frac{\partial c(x,t)}{\partial t} = \frac{\partial}{\partial x}\left(D\frac{\partial c(x,t)}{\partial x}\right). \qquad (3)$$

When diffusion coefficient $D$ can be considered reasonably constant, the second member of (3) is just diffusion coefficient times the second derivative of concentration. A similar diffusion equation can be written for the flux $J(x,t)$. If the concentration function is such that derivative for space and time can be interchanged than, because of (3) and (1), the time derivative of Flux for constant $D$ is:

$$\frac{\partial J(x,t)}{\partial t} = \frac{\partial}{\partial t}\left(-D\frac{\partial c(x,t)}{\partial x}\right) = -D\frac{\partial}{\partial x}\left(\frac{\partial c(x,t)}{\partial t}\right) = -D\frac{\partial}{\partial x}\left(D\frac{\partial^2 c(x,t)}{\partial x^2}\right) =$$
$$= -D\frac{\partial^2}{\partial x^2}\left(D\frac{\partial c(x,t)}{\partial x}\right) = D\frac{\partial^2}{\partial x^2}(J(x,t)) \qquad (4)$$

From (4) it is clear that, for flux function, we have the same differential equation that we have for concentration. If flux is known than concentration can be calculated from (1) by integration:

$$c(x,t) = c(\infty,t) + \frac{1}{D}\int_x^\infty J(x,t)dx . \qquad (5)$$

The usual condition for the value at infinite of concentration is $c(\infty,t)=0$.

**II.1 – General solution for the infinite body.**

We deal here with the diffusion equation for the diffusing elements concentration $c(x,t)$ with constant diffusion coefficient $D$, written in the time and mono-dimensional space domains $(t,x,)$ where $0 \leq t \leq \infty$ and $-\infty \leq x \leq +\infty$.

$$\frac{\partial c(x,t)}{\partial t} = D\frac{\partial^2 c(x,t)}{\partial x^2} . \qquad (6)$$

When the space variable domain is the full real axis we will name it infinite body domain. The solution to equation (6) exists and it is unique providing that we set initial and boundary conditions. Let's define the initial condition:

$$c(x,0) = f(x) , \qquad (7)$$

A solution to this problem has been indicated in the fundamental treatise of Carslaw and Jaeger[5]:

$$c(x,t) = \frac{1}{2\sqrt{\pi Dt}}\int_{-\infty}^{+\infty} e^{-\frac{(x-y)^2}{4Dt}} f(y)dy . \qquad (8)$$

A heuristic derivation of this result can be based on the general solution of (6) by the separation of variables:

$$c(x,t) = X(x)T(t) . \qquad (9)$$

Inserting position (9) in (6) leads to:

$$X(x)\frac{dT(t)}{dt} = DT(t)\frac{d^2 X(x)}{dx^2} \quad , \tag{10}$$

that can be split in two equations:

$$\frac{1}{DT(t)}\frac{dT(t)}{dt} = -\lambda^2 \quad , \tag{11a}$$

$$\frac{1}{X(x)}\frac{d^2 X(x)}{dx^2} = -\lambda^2 \quad , \tag{11b}$$

with general solutions:

$$T(t) = \gamma e^{-\lambda^2 Dt} \quad , \tag{12a}$$

$$X(x) = \alpha \cos(\lambda x) + \beta \sin(\lambda x) \quad . \tag{12b}$$

According to (9) we can write the generic solution:

$$c(x,t) = e^{-\lambda^2 Dt}\left[A(\lambda)\cos(\lambda x) + B(\lambda)\sin(\lambda x)\right] \quad . \tag{13}$$

Because of the linearity of equation (6) the general solution is the one obtained by the linear superposition of the (13) for all possible $\lambda$ values generated by the boundary conditions on the considered domain. The values of $\lambda$ depend on the boundary conditions that will be used, in particular for bounded domains the values of $\lambda$ are discrete, leading to:

$$c(x,t) = \sum_{j=1}^{\infty} e^{-\lambda_j^2 Dt}\left[A(\lambda_j)\cos(\lambda_j x) + B(\lambda_j)\sin(\lambda_j x)\right], \tag{14}$$

while for unbounded domains they are continuous leading to:

$$c(x,t) = \int_{-\infty}^{+\infty} e^{-\lambda^2 Dt}\left[A(\lambda)\cos(\lambda x) + B(\lambda)\sin(\lambda x)\right]d\lambda \quad . \tag{15}$$

We consider here solution (15) because we have an unbounded domain. $A(\lambda)$ and $B(\lambda)$ functions can be determined by considering the initial condition (7):

$$c(x,0) = f(x) = \int_{-\infty}^{+\infty}\left[A(\lambda)\cos(\lambda x) + B(\lambda)\sin(\lambda x)\right]d\lambda \quad . \tag{16}$$

We can express the function $f(x)$ considering the Fourier theorem[6]:

$$f(x) = \frac{1}{2\pi}\int_{-\infty}^{+\infty}\left[\int_{-\infty}^{+\infty} f(\eta)\cos\lambda(\eta - x)d\eta\right]d\lambda =$$

$$= \int_{-\infty}^{+\infty}\left\{\left[\frac{1}{2\pi}\int_{-\infty}^{+\infty} f(\eta)\cos(\lambda\eta)d\eta\right]\cos(\lambda x) + \left[\frac{1}{2\pi}\int_{-\infty}^{+\infty} f(\eta)\sin(\lambda\eta)d\eta\right]\sin(\lambda x)\right\}d\lambda \tag{17}$$

comparing equations (17) and (16) we have:

$$A(\lambda) = \frac{1}{2\pi} \int_{-\infty}^{+\infty} f(\eta)\cos(\lambda\eta)d\eta \quad , \tag{18a}$$

$$B(\lambda) = \frac{1}{2\pi} \int_{-\infty}^{+\infty} f(\eta)\sin(\lambda\eta)d\eta \quad . \tag{18b}$$

From (15) and (18a) and (18b) the general solution is:

$$c(x,t) = \frac{1}{2\pi} \int_{-\infty}^{+\infty} e^{-\lambda^2 Dt} \left[ \int_{-\infty}^{+\infty} f(\eta)\left[\cos(\lambda\eta)\cos(\lambda x) + \sin(\lambda\eta)\sin(\lambda x)\right]d\eta \right] d\lambda =$$

$$= \frac{1}{2\pi} \int_{-\infty}^{+\infty} f(\eta)\left[ \int_{-\infty}^{+\infty} e^{-\lambda^2 Dt} \cos\lambda(\eta - x)d\lambda \right] d\eta \tag{19}$$

The integral in square brackets is from Budak and Fomin[6] page 573:

$$\int_{-\infty}^{+\infty} e^{-\lambda^2 Dt} \cos\lambda(\eta - x)d\lambda = \sqrt{\frac{\pi}{Dt}} e^{\left[-\frac{(\eta-x)^2}{4Dt}\right]} \quad . \tag{20}$$

After (20), the solution for concentration (19) results :

$$c(x,t) = \frac{1}{2\sqrt{\pi Dt}} \int_{-\infty}^{+\infty} f(\eta) e^{-\frac{(\eta-x)^2}{4Dt}} d\eta \quad . \tag{21}$$

Equation (21) is the same of equation (8). The same result, namely equations (8) and (21), is achieved in a number of different ways as indicated also in Carslaw and Jaeger[5], we prefer this heuristic derivation to stress connections to boundary value problems in either bounded and unbounded domains.

**II.2 General solution for the semi-infinite body**

In many scientific and technological problems of interest in materials science we can assume, with a good approximations, that the solid body under diffusion of externally delivered elements or impurities is a semi-infinite body extended in the $x \geq 0$ spatial dimension. In this case we can conveniently use the solution for the infinite body (16) by extending the initial condition towards the negative half space assuming an indeterminate function $c_1(x,0)$ $(x<0)$ and defining the following initial condition:

$$f(\eta) = c(\eta,0) \; ; \; \eta > 0 \quad , \tag{22a}$$

$$f(\eta) = c_1(\eta,0) \; ; \; \eta < 0 \quad . \tag{22b}$$

With these positions solution (21) results:

$$c(x,t) = \frac{1}{2\sqrt{\pi Dt}}\left[\int_{-\infty}^{0} c_1(\eta,0)e^{-\frac{(\eta-x)^2}{4Dt}} d\eta + \int_{0}^{\infty} c(\eta,0)e^{-\frac{(\eta-x)^2}{4Dt}} d\eta\right] \quad . \tag{23}$$

Taking the first integral in (23) and putting $y=-\eta$ than:

$$\int_{-\infty}^{0} c_1(\eta,0)e^{-\frac{(\eta-x)^2}{4Dt}} d\eta =$$

$$= -\int_{+\infty}^{0} c_1(-y,0)e^{-\frac{(-y-x)^2}{4Dt}} dy = \int_{0}^{\infty} c_1(-y,0)e^{-\frac{(x+y)^2}{4Dt}} dy = \int_{0}^{\infty} c_1(-\eta,0)e^{-\frac{(x+\eta)^2}{4Dt}} d\eta \quad , \tag{24}$$

Hence:

$$c(x,t) = \frac{1}{2\sqrt{\pi Dt}}\left[\int_{0}^{\infty} c_1(-\eta,0)e^{-\frac{(\eta+x)^2}{4Dt}} d\eta + \int_{0}^{\infty} c(\eta,0)e^{-\frac{(\eta-x)^2}{4Dt}} d\eta\right] =$$

$$= \frac{1}{2\sqrt{\pi Dt}}\int_{0}^{\infty}\left[c(\eta,0)e^{-\frac{(\eta-x)^2}{4Dt}} + c_1(-\eta,0)e^{-\frac{(\eta+x)^2}{4Dt}}\right] d\eta \quad . \tag{25}$$

**II.3 Semi-infinite body with reflecting boundary (SI-RB)**

Reflecting boundary condition for the semi-infinite body means that no flow of matter is assumed at $x=0$. This can be conveniently defined by using the flux equation and putting it equal to zero at the half plane interface:

$$J(x=0,t) = -D\left.\frac{\partial c}{\partial x}\right|_{x=0} = 0 \quad . \tag{26}$$

Boundary condition (22) can be introduced by differentiating (25):

$$\frac{\partial c(x,t)}{\partial x} = \frac{1}{2Dt}\frac{1}{2\sqrt{\pi Dt}}\int_{0}^{\infty}\left[(x-\eta)c(\eta,0)e^{-\frac{(\eta-x)^2}{4Dt}} d\eta - (\eta+x)c_1(-\eta,0)e^{-\frac{(x+\eta)^2}{4Dt}}\right] d\eta \quad , \tag{27}$$

and putting this to zero (26):

$$\left.\frac{\partial c(x,t)}{\partial x}\right|_{x=0} = \frac{1}{2Dt}\frac{1}{2\sqrt{\pi Dt}}\int_{0}^{\infty}[c(\eta,0) - c_1(-\eta,0)]\eta e^{-\frac{\eta^2}{4Dt}} d\eta = 0 \quad . \tag{28}$$

Condition (28) is satisfied only if:

$$c(\eta,0) = c_1(-\eta,0) \quad .\tag{29}$$

Condition (29) in (25) leads to the general solution for the SI-RB problem:

$$c(x,t) = \frac{1}{2\sqrt{\pi Dt}} \int_0^\infty c(\eta,0) \left[ e^{-\frac{(\eta-x)^2}{4Dt}} + e^{-\frac{(\eta+x)^2}{4Dt}} \right] d\eta \quad .\tag{30}$$

**II.4 Semi-infinite body with capturing boundary (SI-CB)**

The capturing boundary condition is expressed as follows:

$$c(0,t) = 0 \quad .\tag{31}$$

This condition in (25) leads to:

$$c(0,t) = \frac{1}{2\sqrt{\pi Dt}} \int_0^\infty [c(\eta,0) + c_1(-\eta,0)] e^{-\frac{(\eta)^2}{4Dt}} d\eta \quad ,\tag{32}$$

hence:

$$c(\eta,0) = -c_1(-\eta,0),\tag{33}$$

and the general solution for the semi infinite capturing boundary (SI-CB) is:

$$c(x,t) = \frac{1}{2\sqrt{\pi Dt}} \int_0^\infty c(\eta,0) \left[ e^{-\frac{(\eta-x)^2}{4Dt}} - e^{-\frac{(\eta+x)^2}{4Dt}} \right] d\eta \quad .\tag{34}$$

A particular application of solution (34) can be considered for the following initial and boundary conditions:

$$c(x,0) = 0 \tag{35a}$$

$$c(x,t)\big|_{x=0} = C_S \tag{35b}$$

Let's introduce the auxiliary function:

$$c^*(x,t) = C_s - c(x,t) \tag{36}$$

The diffusion problem for the auxiliary function with conditions (35a) and (35b) is:

$$\frac{\partial c^*(x,t)}{\partial t} = D \frac{\partial^2 c^*(x,t)}{\partial x^2} ; \tag{37a}$$

$$c^*(0,t) = 0. \tag{37b}$$

Solution to (37a) and (37b) is the one for the SI-CB

$$c^*(x,t) = \frac{1}{2\sqrt{\pi Dt}} \int_0^\infty c^*(\eta,0) \left[ e^{-\frac{(\eta-x)^2}{4Dt}} - e^{-\frac{(\eta+x)^2}{4Dt}} \right] d\eta \tag{38}$$

$$c^*(\eta,0) = C_s - c(\eta,0) = C_s \tag{39}$$

and:

$$c^*(x,t) = \frac{C_s}{2\sqrt{\pi Dt}} \int_0^\infty \left[ e^{-\frac{(\eta-x)^2}{4Dt}} - e^{-\frac{(\eta+x)^2}{4Dt}} \right] d\eta = C_s erf\left(\frac{x}{2\sqrt{Dt}}\right), \tag{40}$$

The justification of (40) is simple but not trivial hence, following the way of detailed proofing we provide a justification in Appendix (1). After (40) we finally have the so called "erfc" solution:

$$c(x,t) = C_s - c^*(x,t) = C_S\left(1 - erf\left(\frac{x}{2\sqrt{Dt}}\right)\right) = C_s erfc\left(\frac{x}{2\sqrt{Dt}}\right) \quad . \tag{41}$$

Solution (41) is a well-known and, in some ways, popular solution in the theory of diffusion in solids in a semi-infinite body[5,7,8,9].

### III. A variable boundary condition diffusion problem

Now we deal with a diffusion problem with a variable boundary condition. Such type of problems are often encountered when diffusion species are introduced and let diffuse in a solid body by different sequential mechanisms like for example diffusion from a continuous source like a bath or a vapor environment than stopped and followed by a thermal treatment at different temperature. We are typically in front of a two steps diffusion process characterized by different boundary conditions for each step. They can be modelled as constant source up to a certain time than changed as a limited source diffusing in the body without incoming flux through the surface. In this case our problem can be split into two steps separated boundary value problems for the diffusion equation (6):

$$c(0,t) = C_s \text{ for } 0 \le t \le \tau \text{ with a diffusion coefficient } D_0 \quad , \tag{42}$$

$$\left.\frac{\partial c(x,t)}{\partial x}\right|_{x=0} = 0, \text{ for } t \ge t \text{ with a diffusion coefficient } D \quad . \tag{43}$$

The solution up to t=τ is easily found in equation (41) that, at t=τ, becomes itself the initial condition for problem with boundary condition (43):

$$c(x,\tau) = C_s erfc\left(\frac{x}{2\sqrt{D_0\tau}}\right) \quad . \tag{44}$$

Let's rewrite problem (6), (43) and (44) in terms of flux:

$$\frac{\partial J(x,t)}{\partial t} = D\frac{\partial^2}{\partial x^2}(J(x,t)) \qquad (45)$$

$$J(0,t) = 0 \qquad (46)$$

Looking to (45) and (46) we recognize a SIB-CB problem for the flux function $J(x,t)$ and the solution reads (38):

$$J(x,t) = \frac{1}{2\sqrt{\pi Dt}}\int_0^\infty J(\eta,0)\left[e^{-\frac{(\eta-x)^2}{4Dt}} - e^{-\frac{(\eta+x)^2}{4Dt}}\right]d\eta =$$

$$\frac{e^{-\frac{x^2}{4Dt}}}{2\sqrt{\pi Dt}}\int_0^\infty J(\eta,0)e^{-\frac{(\eta)^2}{4Dt}}\left[\sinh\left(\frac{x\eta}{2Dt}\right)\right]d\eta \qquad (47)$$

This is obtained by the development of the squares in the arguments of the exponentials and because $\sinh(z) = \frac{1}{2}(e^z - e^{-z})$. Additionally, the initial condition for the flux is:

$$J(\eta,0) = -D\left.\frac{\partial c(\eta,t)}{\partial \eta}\right|_{t=0} = C_s D\sqrt{\frac{1}{D_0\pi\tau}}e^{-\eta^2/4D_0\tau} \qquad (48)$$

finally the solution for the flux function reads:

$$J(x,t) = C_s\sqrt{\frac{D^2}{D_0\pi\tau}}\frac{e^{-\frac{x^2}{4Dt}}}{\sqrt{\pi Dt}}\int_0^\infty e^{-\eta^2\left(\frac{1}{4D_0\tau}+\frac{1}{4Dt}\right)}\left[\sinh\left(\frac{x\eta}{2Dt}\right)\right]d\eta =$$

$$= \frac{C_s}{\pi\sqrt{t\tau}}\sqrt{\frac{D}{D_0}}e^{-\frac{x^2}{4Dt}}\int_0^\infty e^{-\eta^2\left(\frac{Dt+D_0\tau}{4DD_0 t\tau}\right)}\left[\sinh\left(\frac{x\eta}{2Dt}\right)\right]d\eta \qquad (49)$$

Let's make the following positions:

$$a = \frac{Dt + D_0\tau}{4DD_0 t\tau}; b = -\frac{x}{4Dt}$$

$$\frac{b}{\sqrt{a}} = -\frac{x}{4Dt}\sqrt{\frac{4DD_0 t\tau}{Dt + D_0\tau}} = -\frac{x\sqrt{D_0\tau}}{2\sqrt{Dt(Dt + D_0\tau)}} \qquad (50)$$

$$\frac{b^2}{a} = \frac{x^2 D_0\tau}{4Dt(Dt + D_0\tau)}$$

with these positions the solution for flux (49) results:

$$J(x,t) = \frac{C_s}{\pi\sqrt{t\tau}} \sqrt{\frac{D}{D_0}} e^{-\frac{x^2}{4Dt}} \int_0^\infty e^{-a\eta^2} \frac{1}{2}\left[e^{-2b\eta} - e^{2b\eta}\right] d\eta \qquad (51)$$

The integral term in (51) can be evaluated as follows:

$$I(a,b) = \frac{1}{2}\int_0^\infty e^{-a\eta^2}\left[e^{-2b\eta} - e^{2b\eta}\right] d\eta = \frac{1}{4}\sqrt{\frac{\pi}{a}} e^{\frac{b^2}{a}} erf\left(\frac{b}{\sqrt{a}}\right) \qquad (52)$$

This is coming from Abramowitz and Stegun[10] 7.4.2:

$$\int_0^\infty e^{-(a\eta^2 + 2b\eta + c)} d\eta = \frac{1}{2}\sqrt{\frac{\pi}{a}} e^{\frac{b^2 - ac}{a}} erf\left(\frac{b}{\sqrt{a}}\right) \qquad , \qquad (53)$$

and from the complementary error function *erfc(x)=1-erf(x)* properties such that:

$$erfc(z) - erfc(-z) = -2erf(z) \qquad . \qquad (54)$$

Developing (52) with positions (50) results:

$$I(a,b) = \sqrt{\frac{4\pi D D_0 t\tau}{Dt + D_0\tau}} erf\left(\frac{x\sqrt{D_0\tau}}{2\sqrt{Dt(Dt + D_0\tau)}}\right) e^{\frac{x^2 D_0\tau}{4Dt(Dt + D_0\tau)}} \qquad . \qquad (55)$$

Finally flux solution (51) is:

$$J(x,t) = \frac{C_s}{\pi\sqrt{t\tau}}\sqrt{\frac{D}{D_0}}\sqrt{\frac{\pi D D_0 t\tau}{Dt + D_0\tau}} e^{-\frac{x^2}{4Dt}\left(1 - \frac{D_0\tau}{Dt + D_0\tau}\right)} erf\left(\frac{x\sqrt{D_0\tau}}{2\sqrt{Dt(Dt + D_0\tau)}}\right) =$$

$$= C_s \sqrt{\frac{D^2}{\pi(Dt + D_0\tau)}} e^{-\frac{x^2}{4(Dt + D_0\tau)}} erf\left(\frac{x\sqrt{D_0\tau}}{2\sqrt{Dt(Dt + D_0\tau)}}\right) \qquad . \qquad (56)$$

From the flux solution represented by equation (56) we can obtain the corresponding concentration solution by equation (5) that is integrating equation (56). Let's fix the following positions:

$$\gamma^2 = 4(D_0\tau + Dt); k^2 = \frac{D_0\tau}{Dt}$$

$$x' = \frac{x}{\gamma}; dx = \gamma dx' = 2\sqrt{D_0\tau + Dt}\, dx' \qquad , \qquad (57)$$

$$x'^2 = \frac{x^2}{4(Dt + D_0\tau)}; kx' = \sqrt{\frac{D_0\tau}{Dt}} \frac{x}{2\sqrt{D_0\tau + Dt}}$$

integrating (56) in $x$ according to (5), because $c(\infty,t)=0$, considering positions (57) for change in variables it results:

$$c(x,t) = 2\frac{C_s}{\sqrt{\pi}} \int_{x/\gamma}^{\infty} e^{-x'^2} erf(kx')dx' \qquad (58)$$

$x'$ is an integration variable that we can change in $y$ so we write the final solution:

$$c(x,t) = 2\frac{C_s}{\sqrt{\pi}} \int_{x/\gamma}^{\infty} e^{-y^2} erf(ky)dy;$$

$$\gamma = 2\sqrt{D_0\tau + Dt}; k = \sqrt{\frac{D_0\tau}{Dt}} \qquad (59)$$

where $D_0$ is the diffusion coefficient of the first step diffusion problem (42) up to time $t=\tau$ and $D$ is the diffusion coefficient of problem (43) up to time $t$.

The solution to the two-steps diffusion problem can be finally summarized as follows:

$$c(x,t) = C_s erfc\left(\frac{x}{2\sqrt{D_0 t}}\right); \qquad 0 \le t \le \tau \qquad (60)$$

$$c(x,t) = 2\frac{C_s}{\sqrt{\pi}} \int_{x/\gamma}^{\infty} e^{-y^2} erf(ky)dy;$$

$$\gamma = 2\sqrt{D_0\tau + Dt}; k = \sqrt{\frac{D_0\tau}{Dt}} \qquad \tau \le t < \infty$$

**IV Discussion of the solution for the two step diffusion problem**

The exact solution for the two steps diffusion problem (equation (60) can be discussed in two limiting cases. In order to go ahead with this analysis let's put the following definitions:

$$t' = \tau + t \qquad (61)$$

Indicating with $\tau$ the time of the first diffusion problem step and with $t$ the time of the second diffusion step and with $t'$ the overall (Step1+Step2) diffusion time. When $t\rightarrow 0$, that is when $t'\rightarrow \tau$, than $k\rightarrow\infty$ and $erf(ky)=1$. Under these limiting conditions $\gamma \simeq 2\sqrt{D_0\tau}$ and the solution (60) for the second diffusion step reads:

$$c(x,t) = 2\frac{C_s}{\sqrt{\pi}} \int_{x/\sqrt{D_0 t}}^{\infty} e^{-y^2} dy = C_s erfc(\frac{x}{D_0 t}) \qquad (62)$$

As expected, when $t' \approx \tau$ we can use the constant source solution that is the solution for the first step diffusion problem. More interesting is the case for large times of the second diffusion step that is $t \to \infty$. To analyze this condition we need to manipulate the second step solution of equation (60) taking advantage of the integral transformation (see Malkovich[3]):

$$I = \int_n^\infty e^{-py^2} erf(qy) dy = \frac{1}{\sqrt{\pi}} e^{-p^2 n^2} \int_0^q \frac{e^{-z^2 n^2}}{p^2 + z^2} dz \tag{63}$$

The transformation is obtained considering first integral (63) with upper limit $n$ than differentiating through q and finally taken the limit to infinite. According to (63) the solution (60) resukts:

$$c(x,t) = 2\frac{C_s}{\sqrt{\pi}} \int_{x/\gamma}^\infty e^{-y^2} erf(ky) dy = 2\frac{C_s}{\sqrt{\pi}} \frac{e^{-\left(\frac{x}{\gamma}\right)^2}}{\sqrt{\pi}} \int_0^k \frac{e^{-\left(\frac{x}{\gamma}\right)^2 z^2}}{1+z^2} dz \tag{64}$$

In the limiting condition $t >> \tau$, $k$ is expected to be close to zero and we can assume:

$$\int_0^k f(x,t,z) dz \approx k \cdot f(x,t,0) \tag{65}$$

But for $z \to 0$ the function under the integral in (64) is 1, hence the limiting solution for $t >> \tau$ is:

$$c(x,t) \approx 2\frac{C_s}{\pi} \sqrt{\frac{D_0 \tau}{Dt}} e^{-\left(\frac{x^2}{4Dt}\right)} \tag{66}$$

The total quantity of diffusion elements entered into the solid body during the first diffusion step is easily calculated integrating the concentration from time $0$ to $t$ leading to a well known result (see Crank[7] page 32):

$$Q(\tau) = 2C_s \sqrt{\frac{D_0 \tau}{\pi}} \tag{67}$$

We can express the limiting solution (66) in terms of the total quantity of diffusing elements entered in step 1 (67) and it results:

$$c(x,t) \approx \frac{Q}{\sqrt{\pi Dt}} e^{-\left(\frac{x^2}{4Dt}\right)} . \tag{68}$$

It is nice to realize that equation (68) is exactly the concentration solution for the diffusion in a semi-infinite body for the diffusion of a thin layer (see again Crank[7] page 13).

Summarizing, we can define the following conditions:

If $t' \approx \tau$ we can use the constant source solution of step 1:

$$c(x,t) = C_s \, erfc(\frac{x}{D_0 t}) \tag{i}$$

If $t \gg \tau$ (say $t > 4\tau$) we can use the infinitesimal thin layer solution:

$$c(x,t) \approx \frac{Q}{\sqrt{\pi D t}} e^{-\left(\frac{x^2}{4Dt}\right)} \tag{ii}$$

In between these times: $\tau < t < 4\tau$, we shall use the exact solution:

$$c(x,t) = 2\frac{C_s}{\sqrt{\pi}} \int_{x/\gamma}^{\infty} e^{-y^2} erf(ky)dy;$$

$$\gamma = 2\sqrt{D_0 \tau + Dt}; k = \sqrt{\frac{D_0 \tau}{Dt}} \tag{iii}$$

**Appendix 1. Justification of equation (40)**

Equation (40) reads:

$$c^*(x,t) = \frac{C_s}{2\sqrt{\pi Dt}} \int_0^\infty \left[ e^{-\frac{(\eta-x)^2}{4Dt}} - e^{-\frac{(\eta+x)^2}{4Dt}} \right] d\eta = C_s \, erf\left(\frac{x}{2\sqrt{Dt}}\right) \quad \text{(A1)}$$

To justify this result let's consider the integrals:

$$\int_0^\infty \left[ e^{-\frac{(\eta-x)^2}{4Dt}} - e^{-\frac{(\eta+x)^2}{4Dt}} \right] d\eta = \int_0^\infty e^{-\frac{(\eta-x)^2}{4Dt}} d\eta - \int_0^\infty e^{-\frac{(\eta+x)^2}{4Dt}} d\eta \quad \text{(A2)}$$

After the change of variables:

$$z = \frac{x-\eta}{\sqrt{4Dt}}; \; dz = -\frac{d\eta}{\sqrt{4Dt}}$$
$$\eta = x - z\sqrt{4Dt} \quad \text{(A3)}$$

The first integral in (A2) can be evaluated:

$$\int_0^\infty e^{-\frac{(\eta-x)^2}{4Dt}} d\eta = -\sqrt{4Dt} \int_{x/\sqrt{4Dt}}^{-\infty} e^{-z^2} dz = \sqrt{4Dt} \int_{-\infty}^{x/\sqrt{4Dt}} e^{-z^2} dz = \sqrt{4Dt} \left[ \int_{-\infty}^{0} e^{-z^2} dz + \int_{0}^{x/\sqrt{4Dt}} e^{-z^2} dz \right]$$

The second integral is evaluated according to the following change of variables:

$$z = \frac{x+\eta}{\sqrt{4Dt}}; \; dz = \frac{d\eta}{\sqrt{4Dt}}$$
$$\eta = x + z\sqrt{4Dt} \quad \text{(A4)}$$

The second integral in (A2) can be evaluated:

$$\int_0^\infty e^{-\frac{(\eta+x)^2}{4Dt}} d\eta = \sqrt{4Dt} \int_{x/\sqrt{4Dt}}^{\infty} e^{-z^2} dz = -\sqrt{4Dt} \int_{\infty}^{x/\sqrt{4Dt}} e^{-z^2} dz = -\sqrt{4Dt} \left[ \int_{\infty}^{0} e^{-z^2} dz + \int_{0}^{x/\sqrt{4Dt}} e^{-z^2} dz \right]$$

The integrals in (A2) can be evaluated as follows:

$$\int_0^\infty \left[ e^{-\frac{(\eta-x)^2}{4Dt}} - e^{-\frac{(\eta+x)^2}{4Dt}} \right] d\eta = \sqrt{4Dt} \left[ \int_{-\infty}^0 e^{-z^2} dz + \int_0^{x/\sqrt{4Dt}} e^{-z^2} dz \right] + \sqrt{4Dt} \left[ \int_\infty^0 e^{-z^2} dz + \int_0^{x/\sqrt{4Dt}} e^{-z^2} dz \right] =$$

$$= 2\sqrt{Dt} \left[ 2 \int_0^{x/\sqrt{4Dt}} e^{-z^2} dz \right] = 2\sqrt{Dt} \sqrt{\pi}\, erf\left(\frac{x}{2\sqrt{Dt}}\right)$$

That proofs equation (A1) and so equation (40).